\documentclass{article}
\usepackage{amsfonts}
\usepackage[dvips]{graphicx}
\usepackage{color} 
\begin{document}
\newcommand\eq[1] {(\ref{#1})}
\title{Optimal Three-Material  Wheel Assemblage of Conducting and Elastic Composites}
\author{Andrej Cherkaev
\\Department of Mathematics, University of Utah }
\maketitle

\abstract{We describe a new type of three material  microstructures which we call {\em wheel assemblages}, 
that correspond to extremal conductivity and extremal bulk modulus for a composite made of two materials and
an ideal material.  The exact lower bounds for effective conductivity and matching laminates 
was 
found in (Cherkaev, 2009) and for anisotropic composites, in (Cherkaev, Zhang, 2011). 
Here, we show different optimal structures that generalize 
the classical Hashin-Shtrikman coated spheres (circles). They consist
of circular inclusions which contain a solid central circle (hub) and radial spikes in a surrounding annulus, 
and (for larger volume fractions of the best material) an annulus filled with it. The same wheel assemblages are optimal for the 
pair of dual problems of minimal conductivity (resistivity) of a composite made from two materials and an ideal conductor (insulator), 
in the problem of maximal effective bulk modulus of elastic composites made from two linear elastic material
and void, and the dual minimum problem. 


\section{Introduction}

This paper  introduces a new type of optimal isotropic structure for multimaterial composites: the wheels. They remain Hashin-Shtrikman coated spheres but consist of not only concentric annuli, but also 
an annulus with spikes: radial  layers that alternate infinitely fast. 
We show optimality of wheel assemblages for the simplest structure of this sort:  Three-phase two-dimensional composites of optimal conductivity, in which one of the phases is either an ideal conductor or a perfectly insulating material.  

Here, we briefly summarize the development of optimal structures. More detailed descriptions can be found in \cite{mybook,miltonbook,cherk09,cz11}. In the pioneering work  \cite{hs63}, Hashin and Shtrikman found the bounds and the matching structures for optimal isotropic two-component composites, and suggested bounds for multicomponent ones. Milton \cite{milton81} showed that the Hashin-Shtikman bound is not exact everywhere (it tends to an incorrect limit when $m_1\to 0$), but is exact when $m_1$ is large enough, $m_1\geq g_m$. Nesi \cite{nesi}  suggested a new tighter bound for isotropic multicomponent structures, and Cherkaev \cite{cherk09} further improved it and found optimal laminates that realize the bounds. These latter two bounds coincide in the case $k_3=\infty$ that is considered here.  The method for 
obtaining bounds is based on the procedure suggested by Nesi \cite{nesi} that combines the translation method \cite{mybook,miltonbook} and additional inequality constraints \cite{nesi-ineq}. Cherkaev and Zhang \cite{cz11} derive bounds for anisotropic composites made from two isotropic materials and void. 

The geometry of optimal isotropic two--material micro-geometries is intuitively clear:  the stronger material ``wraps" the weaker one so that the weak material forms a nucleus, and the strong one - an envelope.  
This principle is reflected in the first ``coated spheres" geometry of optimal structures found by Hashin and Shtrikman,  and in following development: Lurie and Cherkaev \cite{luriecherk82,luriecherk86} suggested optimal high-rank laminates, Vigdergauz \cite{vigder}, Grabovsky and Kohn \cite{Grab-kohn} and recently  Lui \cite{lui08} suggested special convex oval-shaped inclusions, Milton \cite{miltonbook} introduced a general method of finding optimal shapes of inclusions, and recent paper by Benveniste and Milton \cite{benveniste} investigated "coated ellipsoids".
For multimaterial case, the picture is much more diverse.
Milton \cite{milton81} introduced  parallel coated spheres, Lurie and Cherkaev suggested multilayer coated circles \cite{luriecherk85},  Gibiansky and Sigmund \cite{gib-sig} suggested "bulk blocks", Albin and Cherkaev proved the optimality of three-material ``haired spheres" \cite{albin-cherk06}. All these structures admit separation of variables when  effective properties are computed.  However, study of these structures has been 
overshadowed by the convenience of multiscale laminates and most recent results have been proven for laminates. Particularly, in \cite{cherk09, cz11}, it is shown  that proper laminates are optimal for the considered here problem and its anisotropic generalization, respectively. 
 
The isotropic structures - {\em wheels} described here are direct generalizations of Hashin-Shtrikman coated spheres. They are intrinsically isotropic and simpler than isotropic laminates, and they nicely illustrate the qualitative properties of optimal structures. The topology of optimal structures depend on prescribed volume fractions of the materials being mixed. They undergo two phase transitions when the volume fraction of the best material decreases. The optimal microstructures are found by the combination of effective medium method and a procedure suggested in \cite{albin-nesi-cherk07}. Similar  "haired sphere" structures are studied in \cite{albin-cherk06} and it is shown that they realize three material Hashin-Shtrikman bounds for a range of parameters.

\paragraph{Acknowledgment} The research is supported by the grant from DMS NSF.

\section{Equations and Notations}

Consider a composite: The materials $k_i$ occupy plane domains $\Omega_i, ~i=1,2,3 \subset R_2$ whose union forms a unit cell  $\Omega$. 
The areas $m_i =\| \Omega_i \| $ of $\Omega_i$, corresponding to prescribed volume fractions of each material are fixed:
 $ m_1+m_2+m_3=1, ~m_i \geq 0$, and there are no other constraints on $\Omega_i$.
The conductivity of the composite is described by a system of equations 
that relate fields $e$ and  currents $j$
\begin{eqnarray}
 \nabla \times e=0, \quad \mbox{or } e=-\nabla u, ~~ \quad  \nabla \cdot j=0 \quad  \mbox{in } \Omega,
\quad  j=k_i e ~~  \mbox{in } \Omega_i  \label{e-diff}
  \end{eqnarray}
where $u\in H_1(\Omega)$ is a scalar potential. The potential $u$ and the normal current $j \cdot n$ 
are continuous at the boundaries $ \partial_{ik} $ between $\Omega_i$ and $\Omega_k$, so
the following conditions hold:
\begin{equation}
\tau \cdot (e_i-e_k)=0, \quad n \cdot (j_i-j_k)=0 \quad \mbox{at } \partial_{ik} ,
\label{bound-cond}
\end{equation}
where $\tau$ and $n$ are the tangent and normal to $\partial_{ik} $, respectively.

In order to determine effective properties of an isotropic composite, we subject it to two homogeneous orthogonal { external  loadings}  $e_{0a}$ and $e_{0b}$ of equal unit magnitude,
$$
\lim_{|x|\to \infty}e_a(x)= e_{0a}=\pmatrix{1 \cr 0} \quad \mbox{and } \lim_{|x|\to \infty}e_b(x)=e_{0b}=\pmatrix{0 \cr 1}
$$
 and calculate the sum of energies. A pair of corresponding conductivity equations (\ref{e-diff}) (denoted with subindices $a$ and $b$) differ only in boundary conditions. 
This pair can be conveniently viewed as a single problem for a vector potential $U=(u_a, u_b)^T$ and $2 \times 2$ matrices $E=(e_a | e_b)$ and $J=(j_a | j_b)$ in a cell $\Omega$:
\begin{eqnarray}
  \nabla \cdot J=0, ~~   E=-\nabla U \mbox{ in } \Omega, \quad J= k_i\, E \mbox{ in } \Omega_i   \quad \lim_{|x|\to \infty}  E(x) = E_0= I .
\label{vars}
\\
E(x) \mbox{ is $\Omega$-periodic}, \quad  \int_{\Omega} E(x) = E_0, \quad  
   E_0= \pmatrix{1 & 0 \cr 0 & 1}
\label{E0-def}  \label{aver}
\end{eqnarray}

\paragraph{Energy}

In each material $k_i$,  the energy density $W_i={W_i}_a + {W_i}_b$  of the pair is defined as a sum of two energies ${W_i}_a$ and ${W_i}_b$, caused by the  loadings $e_{0a}$ and $e_{0b}$, respectively,
\begin{eqnarray*}
 W_i=  {W_i}_a+{W_i}_b= \frac{1}{2} k_i \mbox{Tr} (E\,E^T)\label{11}
\end{eqnarray*}
where $ \mbox{Tr}$ denotes the trace. 

Here, we assume that materials are ordered:
$$
0< k_1\leq k_2 \leq k_3=\infty
$$ 
and $k_3$ is an ideal conductor. In $ \Omega_3 $, field $E$ is zero and the current is not defined. The energy of the ideal conductor is 
\begin{equation} 
W_3(E)=\left\{\begin{array}{ll}
0 & \mbox{if }E=0 \\
+\infty & \mbox{if }E \neq 0 
\end{array} \right. .
\label{en3}
\end{equation}

The energy $W_0(E_0)$ of the whole periodicity cell has the form
$$
W_0(E_0)=\inf_{E {\rm \,as \,in\,} ( \ref{vars}), (\ref{aver}) } \sum_{i=1}^2\int_{ \Omega_i}W_{i}(E) dx,
$$
It defines the effective conductivity $k_*$ through the formula
\begin{equation}
W_0(E)= \frac{k_*}{2}  \mbox{Tr} (E_0\,E_0^T)=k_*
\label{en-def}
\end{equation}
(because $\mbox{Tr} (E_0\,E_0^T)=2$).
Effective conductivity  $k_*$ depends on $k_i, m_i, $ and on subdivision $\Omega_i$.
The geometrically independent lower bound of the effective conductivity $B$ is the solution of the problem
\begin{equation}
B=  \inf_{\Omega_i: ~\|\Omega_i\|=m_i, \cup \Omega_i=\Omega} \left( \inf_{E {\rm \,as \,in\,} ( \ref{vars}), (\ref{aver}) }  \sum_i \int_{ \Omega_i}W_{i}(E)\,dx \right)
\label{bound-def}
\end{equation}
where $m_1$ $m_2$ and $m_3$ are fixed fractions of materials, $m_1+m_2+m_3=1$.

\section{Optimal bounds and optimal fields}

 \subsection{Exact Bounds and optimal fields}
 \paragraph{Lower bound for effective conductivity}
The problem of exact bounds for an isotropic three material composite was solved in \cite{nesi,cherk09} and the problem of bounds for an anisotropic composite in \cite {cz11}. It was found that effective conductivity $k_*$ of a described  two-dimensional isotropic composite 
is bounded from below by the
bound $B$:
 \begin{equation}
 k_* \geq B(m_1, m_2) \mbox{ for all microstructures}
 \label{b1}
 \end{equation}
The bound has the form, see \cite{cherk09}
 \begin{equation}
B(m_1, m_2)= \left\{\begin{array}{ll}
 B1   &\mbox{ if }
 m_{11}\leq m_1 \leq 1,
\\
B2   &\mbox{ if } m_{11}\leq m_1 \leq m_{12},
\\
B3    & \mbox{ if } ~~0 \leq m_1\leq m_{12}.
 \end{array} \right.
\ \label{b2}
 \end{equation}
 where
\begin{eqnarray} 
 B1  &=& -k_1 + \left(\frac{m_1}{2 k_1}+\frac{m_2}{k_1+k_2}
\right)^{-1} 
\label{b3}.
\\
B2    &=& k_2+2\frac{k_1}{m_1}(1-\sqrt{m_2})^2.  
\label{b4}
\\
B3    &=& -k_2 + \left(\frac{m_1}{2 k_1}+\frac{m_2}{2k_2}
\right)^{-1}.
 \label{b5}
 \end{eqnarray}
and the threshold values $m_{11}$ and $m_{12}$ are
 \begin{equation}
m_{11}=\frac{2 k_1}{k_2+k_1}(\sqrt{m_2}-m_2), \quad
m_{12}=\frac{k_1}{k_2}(\sqrt{m_2}-m_2).
 \label{b6}
  \end{equation}
The bound is a continuously differentiable function of $m_1$ and
$m_2$.

\paragraph{Optimal fields}
The bounds obtained correspond to the following fields in each phase \cite{cherk09}: The superconducting phase naturally corresponds to zero field, $e=0$ in $\Omega_3$, and the fields in $k_1$ and $k_2$ are
\begin{itemize}
\item If $m_1\geq m_{11}$ (Hashin-Shtrikman bound), then
\begin{eqnarray}
{\rm Tr\,} E= \frac{1}{k_1} H_1, ~ \det E \geq 0 ~ \mbox{in } \Omega_1, \quad
E= \frac{1}{k_1+k_2} H_1\,  ~ \mbox{in } \Omega_2 ,
\label{con1}
\\
H_1=\left(\frac{m_1}{2 k_1}+ \frac{m_2}{k_1+k_2} \right)^{-1}. \nonumber
\end{eqnarray}

\item If $m_{12} \leq m_1\leq m_{11}$, (Intermediate bound), then
\begin{eqnarray}
{\rm Tr\,} E= 2\frac{1-\sqrt{m_2}}{m_1}, ~ \det E =0 ~ \mbox{in } \Omega_1, \quad
E= \frac{1}{\sqrt{m_2}} I ~ \mbox{in } \Omega_2. 
\label{con2}
\end{eqnarray}

\item If $m_1\leq m_{12}$ (Small $m_1$ bound), then
\begin{eqnarray}
{\rm Tr\,} E= \frac{2}{k_1} H_2, ~ \det E =0 \quad \mbox{in } \Omega_1,
 \quad
{\rm Tr\,} E= \frac{2}{k_2} H_2 ~ \mbox{in } \Omega_2, 
\label{con3}
\\
H_2=\left(\frac{m_1}{k_1}+ \frac{m_2}{k_2} \right)^{-1} . \nonumber
\end{eqnarray}
\end{itemize}
These fields are derived from sufficient  conditions for optimality, but they may or may not obey the differential equation and jump conditions. It is shown  in \cite{cherk09,cz11}, however, that they are optimal: there exist multiscale laminates (see Figure \ref{f:lam} below) that has exactly the same effective properties as predicted by bounds and the fields that satisfy \eq{con1}, \eq{con2}, \eq{con3}. In the present paper, we show different {\em wheels}  structures that have the same effective conductivity, and the fields in them  satisfy the same conditions.

\subsection{Comments on bounds derivation}
Here we briefly comment on the method used to derive bounds, and see  \cite{nesi,cherk09,cz11} for derivations from 
the viewpoint of convexification. 
To derive the bounds, the problem \eq{bound-def} is reformulated  as a multiwell variational problem
\begin{equation}
J=\inf_{E}\int_\Omega F\left(E, \gamma  \right) dx, \quad F\left(E, \gamma  \right)=\min_{i=1,2, 3}  \left[W_i(E)+ \gamma_i  \right], \quad W_i(E)=\frac{k_i}{2} \|E\|^2
\label{bound}
\end{equation}
 subject to 
 \begin{equation} (a) ~ E=\nabla u, ~\mbox{or } \nabla\times E=0, \quad (b)~ \int_\Omega E\, dx=E_0
 \label{constrains} 
 \end{equation}
 where  $\gamma_1$ are Lagrange multipliers that account for constraints $
 \|\Omega_i\|=m_i. $
The nonconvex multiwell ''energy'' $F$ of an optimal composite depends only on curlfree fields $E$, it is derived using an optimality condition which says that the larger conductivity $k$ corresponds to smaller magnitude $\|E\|$ of $E$. 

Translation method \cite{luriecherk82,luriecherk86,mybook,miltonbook} replaces this variational problem with a minorant problem for $E(x)\in L_2(\Omega)$ omitting the differential constraint (a) in \eq{constrains} but adding instead an integral  constraint of quasiaffineness
\begin{equation} 
t \int_\Omega \det \left(E \right) dx= t \det \int_\Omega  \left(E \right) dx,
\label{quasiaff}
\end{equation}
see \cite{Morrey,reshetnyak}, that is satisfied for all gradient $E=\nabla u$ but obviously not for all $E$. Here $t$ is a real parameter (Lagrange multiplier). 
Energy $W(E)$ and the determinant $\det E$ are conveniently expressed through variables $s$ and $d$
$$
s^2= (E_{11}+ E_{22})^2+ (E_{12}- E_{21})^2, \quad d^2=(E_{11}- E_{22})^2+ (E_{12}+ E_{21})^2
$$
as follows:
$$
W(E)=\frac{k}{4}  \left( s^2+ d^2\right), \quad \det(E)= \frac{1}{4}\left(s^2-d^2\right).
$$
The minimization problem \eq{bound} is estimated from below by the problem 
$$ J\geq \max_{t\in R} \min_{s(x): s_0= 1, ~d(x): d_0=0}\left[\int(F_{\mbox{\footnotesize{transl}}}(s(x), d(x), \gamma_i, t) dx - t \det(E_0) \right]$$
for any real $t$. We compute, fixing $t$ 
\begin{eqnarray}
F_{\mbox{\footnotesize{transl}}} &=& \min_{i=1,2,3}\left\{ W^{\mbox{\footnotesize{transl}}}_i \right\}  
\\  
W^{\mbox{\footnotesize{transl}}}_i &=& W_i (E)+ t \det(E) =
\frac{1}{4}\left[ (k_i+t)  s  ^2+ (k_i-t) d  ^2 \right] + \gamma_i
\label{multiwell} 
\end{eqnarray}
where each well $W^{\mbox{\footnotesize{transl}}}_i$   is the translated energy of a material plus the cost $\gamma_i$. The $L_2(\Omega)$-minimizers $S$ and $d$ are free from any differential constraints, and are subject to only integral constraints
$$
\int_\Omega s(x)\, dx= 1, \quad \int_\Omega d(x)\, dx= 0. 
$$
The convex envelope ${\cal C} F_{\mbox{\footnotesize{transl}}}(s,d,t) $ with respect to $s$ and $d$ represents a lower bound of the composite energy for any value of $t$. The translation bound is obtained by maximization of ${\cal C} F_{\mbox{\footnotesize{transl}}}(s,d,t) $ with respect of $t$. Notice that ${\cal C} F_{\mbox{\footnotesize{transl}}}(s,d,t)=-\infty $ if $|t| > k_1$ because $F^{\mbox{\footnotesize{transl}}}_1$ \eq{multiwell} is a quadratic saddle function. The constraint $|t|  \leq k_1$ must hold independently of the volume fraction $m_1$. Because of it, the translation bound is nonexact for small values of $m_1$ and tends to a wrong asymptotic value when $m_1\to 0$, see the discussion in \cite{milton81,cherk09,cz11}.

The localized polyconvexity  method used here for bounds refines the translation method by  accounting for 
an additional pointwise inequality (Alessandrino, Nesi) \cite{nesi-ineq}
\begin{equation}
 \det(E)= s^2-d^2 \geq 0, ~\forall x\in \Omega
 \label{nesi}
 \end{equation}
that states that the Jacobian of two independent potentials does not change its sign in $\Omega$. To account for this constraint, we define the  translated energy as $+\infty$ if the constraint is violated
 \begin{eqnarray}
F_i( s  , d,t  ) &=& \left\{ 
\begin{array}{ll} W^{\mbox{\footnotesize{transl}}}_i   & \mbox{if } s  ^2- d  ^2 \geq 0, \\
+\infty &  \mbox{if } s  ^2- d  ^2 < 0
\end{array}  
 \right., ~ i=1,2
\nonumber \\
F_3( s  , d, t  ) &=& \left\{ 
\begin{array}{ll}
0 & \mbox{if } s  = d  =0, \\
+\infty & \mbox{otherwise}
\end{array}   \right.\nonumber \
\end{eqnarray}
Notice, that the wells $F_i$ are positive and  grow quadratically with $s, d$ for all real values of parameter $t$, but  $F_1$ and $F_2$  become nonconvex functions of $s, d$ when $t >k_1$ and $t>k_2$, respectively.

The convex envelope ${\cal C } F$ of multiwell translated energy-cost function $F$ 
\begin{eqnarray}
F &=& \min \left\{F_1, \,F_2, \, F_3 \right\}  \label{multiwell2} 
\end{eqnarray}
is nonnegative for all $t$. Therefore the constraint $|t| \leq k_1$ is now lifted and the bound $B$ is the maximum of $F(t)$ with respect of $t\in R_1$. The bound is better than that obtained by the equivalent of the translation bound in this context. 

One can show that the bound is optimal when values of $t$ belong to the interval $[k_1, k_2]$. The translation bound corresponds to $t=k_1$.
When $ t> k_1 $, the well $F_1( s  ,  d  , t)$ becomes nonconvex. To address this case, we first find the convex envelope ${\cal C}F_1( s  ,  d  , t)$ of well $F_1$ 
\begin{equation}
{\cal C}F_1( s  , d  , t)=\left\{ 
\begin{array}{ll}
W^{\mbox{\footnotesize{transl}}}_1  & \mbox{if } s  ^2- d  ^2 \geq 0, ~0\le t\le k_1\\
k_1  s  ^2 + \gamma_1 & \mbox{if } s  ^2- d  ^2 \geq 0, ~t \geq  k_1\\
+\infty & \mbox{if } s  ^2- d  ^2 < 0
\end{array}  
\right.
\end{equation} 
The above bounds \eq{b2} and optimal fields are obtained by analyzing the finite-dimensional optimization problem 
of constructing ${\cal C} F$. Three cases correspond to optimal values of $t$ being $t_{opt}=k_1$ (B1), 
$t_{opt} \in(k_1, k_2)$ (B2), $t_{opt}=k_2$ (B3), respectively. 

{\bf Comment} One can show that a point on convex envelope of 
piece-wise quadratic wells corresponds to the paraboloid with a coefficient equal to harmonic mean of the coefficients of translated energies, see bounds B1 and B3. The irrationality in the expression of B2 is due to optimization with respect to translation parameter $t$. The details of 
this calculation are shown in \cite{cherk09, cz11}

\subsection{Effective medium theory}
We show that the bounds above are realizable by a special structures constructed by means of the effective medium theory. 
Consider the Hashin-Shtrikman coated spheres scheme with anisotropic multilayer inclusions (wheels) of the following geometry. Assume that an infinite plane is filled with a material $k_*$ and has a circular inclusion of unit radius. The central part of the inclusion of radius $r_0$, hub, is filled by a material $k_2$. It is surrounded by an annulus $r_0 < r < 1$ with an axisymmetric anisotropic material $K_{int}(r)$ whose properties may vary with $r$ but not with $\theta$. Assume also that the eigendirections of  $K_{int}(r)$  are $r$ and $\theta$. 

Because of axial symmetry of the inclusion, there is no need to consider two orthogonal external loadings: They correspond to the same but rotated solution. Assume also that the potential $U$, gradient field $E$ and current $J$ have the representation 
$$
U= u(r) \cos(\theta), \quad E=e(r) \cos(\theta), \quad J=j(r) \cos(\theta), 
$$
that corresponds to a homogeneous applied field. 

In polar coordinates the conductivity equations have the following form:
Ohm's law
\begin{eqnarray}
j_r=K_{rr}e_r \quad 
J_\theta= K_{\theta\theta}e_\theta ,
\end{eqnarray}
where the field $e=\nabla u$ is represented as
\begin{eqnarray}
e_r=\frac{\partial }{\partial r}u \quad
e_{\theta}=\frac{1}{r}\frac{\partial }{\partial r} u
\end{eqnarray}
and the equilibrium of currents ($\nabla\cdot J=0$)
\begin{equation}
\frac{1}{r}\frac{\partial }{\partial r} \left(r\, j_{r}\right) + \frac{1}{r^2}\frac{\partial }{\partial \theta} j_\theta=0
\end{equation}

The boundary value problem for $u$ is 
 \begin{eqnarray}
L_u (u)=\left(\frac{d }{d r} r \, K_{r} \frac{d }{d r} - \frac{1}{r} K_{\theta \theta} \right) u=0 \quad r <r_{incl} \label{radeq}\\
\left. K_{rr}\frac{ d u}{dr} \right|_{r \to r_{incl}-0}=\left. k_* \frac{ d u}{dr}\right|_{r \to r_{incl}+0} ,  \quad \mbox{at } r=r_{incl} \label{radbc}, \quad 
\left. \frac{ d u}{dr}\right|_{r \to \infty}=1, \quad u(0)=0
\end{eqnarray}
The energy of the assembly is 
\begin{equation}
 W=\frac{1}{2} \int_0^1{  \left[K_{r}\left(\frac{d \, u}{d r}\right)^2 + \frac{K_\theta}{r^2} u^2\right] r \, dr}
\label{en-cond}
\end{equation}

\paragraph{Effective conductivity}
 
The effective conductivity of the assembly is computed using the Hashin-Shtrikman {\em effective medium} scheme, as follows:  The homogeneous linear potential $u_0 x=u_0 r \cos \theta $ is applied at infinity,
\begin{equation}
\lim_{r\to \infty} u(u, \theta)= u_0 \,r \cos \theta
\end{equation}
Outside the inclusion, where $k=k_*$, potential $u(r)$ satisfies Laplace equation  
and is equal to
\begin{equation}
U= u_0\, r + \frac{B}{r}
 \end{equation}
where  $B$ is a perturbation caused by the inclusion. At the boundary of the inclusion, potential $u$ and the normal current $j_n=k \frac{d }{d r} u$ are continuous. 

If $k_*$ is chosen so that $B=0$, the inclusion becomes "invisible" for an outside observer. In this case, $k_*$ is called {\em effective medium} because the whole plane can be filled with identical inclusions at all scales, without changing the potential at the observer's position. 

The effective conductivity can be conveniently computed from the energy of the assembly and comparing it with the energy of the "invisible inclusion" with undisturbed potential $u_0\, r$. 
$$ W_0= \frac{1}{2} k_* \int_0^1 \left[ r \left(\frac{d \, u}{d r} \right)^2 +\frac{u^2}{ r} \right] r \, dr = k_*u_0^2  $$
 Notice that  $k_*$ is independent of magnitude of $u$. Let us set $u_0=1$; then the energy of the inclusion is equal to $k_*$.

\section{Optimal Wheel Assemblages}

\subsection{Wheel assemblages $W(2,13)$. Intermediate regime}
The Wheel  $W(2, 13)$ structure (nucleus from $k_2$ then radial spikes from $k_1$ alternated with trapezoids from $k_3=\infty$), Figure \ref{f:wheel-i} realizes the intermediate bound B2 \eq{b2} when $m_1 \in [m_{11}, m_{12}]$.
\begin{figure}[htbp]
\begin{center}
\includegraphics[scale=.25]{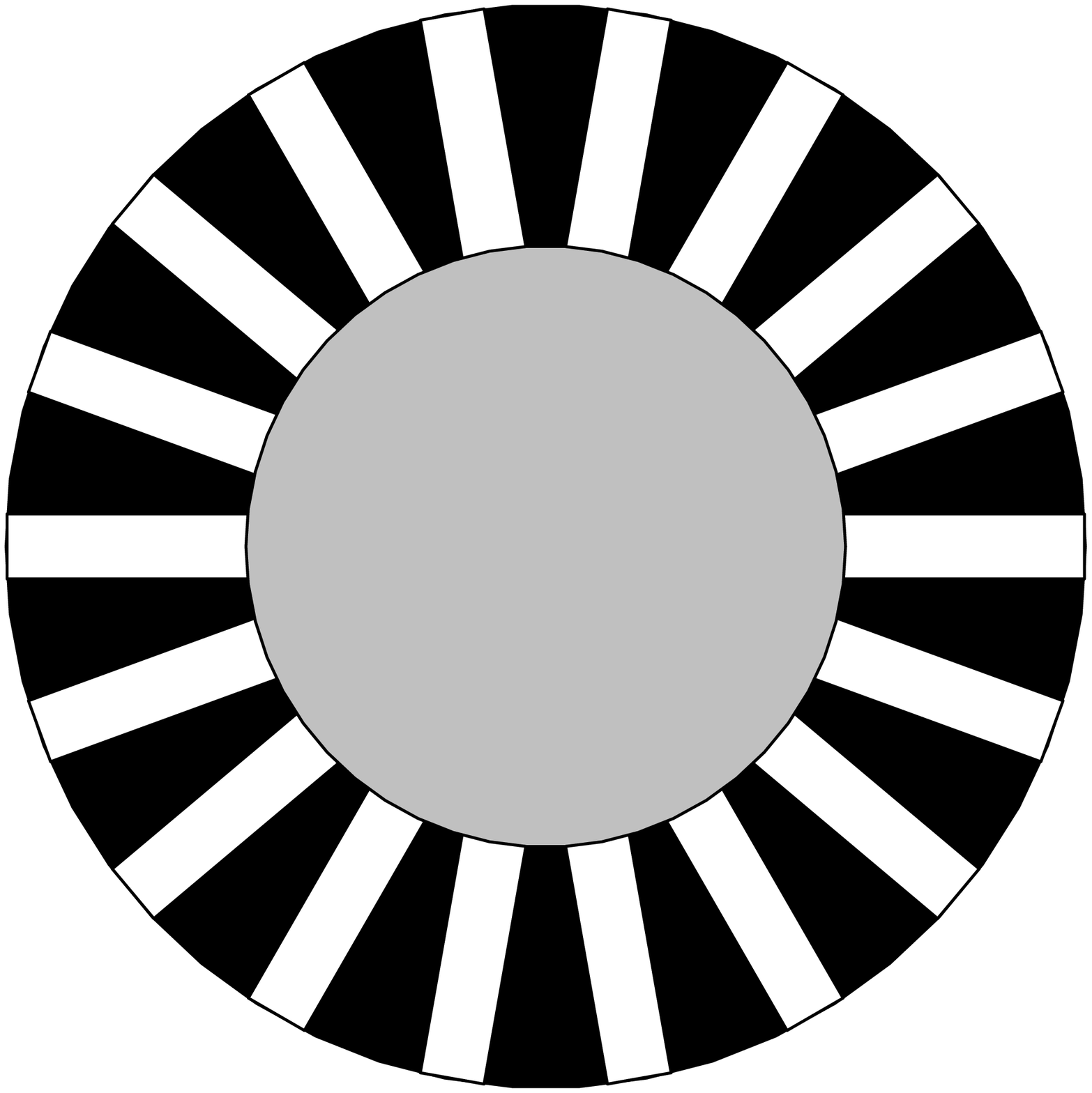} 
\caption{Cartoon of optimal  Wheels $W(2, 13)$. Intermediate case $m_{12} < m_1< m_{11}$. White zones correspond to $k_1$, grey - to $k_2$, and black � to $k_3=\infty$ }
\label{f:wheel-i}
\end{center}
\end{figure}
The geometrical parameters of the structure are as follows:
The central circle $(r<r_0=\sqrt{m_2})$ is occupied by material $k_2$. The outer annulus $\sqrt{m_2}\leq r \leq 1 $ is filled with infinitesimal radial strips of constant width occupied by $k_1$ and trapezoids occupied by $k_3$.  The total thickness of $k_1$-strips is denoted  $b \, r_0$, the total area $b r_0 (1-r_0) $ of them is equal to the fixed fraction $m_1$: 
 $$ m_1= b r_0 (1-r_0)=b\sqrt{m_2}(1-\sqrt{m_2})
 $$
 therefore  $$
 b=\frac{m_1} {\sqrt{m_2}(1-\sqrt{m_2})}.
 $$
 $b$ varies with volume fraction $m_1$, $ b\in \left[b_{11}, \, b_{12} \right]$
where
 $$b_{11}= b|_{m_1=m_{11}}= \frac{2k_1}{k_1+ k_2} \quad   b_{12}= b|_{m_1=m_{12}}= \frac{k_1}{ k_2}.  $$

The potential $u(r)$ is continuous, we set $u(0)=0$, $u(1)=1$. The jump condition for the currents is not considered because $K_{rr}=\infty$ in the middle layer and the current is not defined. We compute
$$
u(r)=\left\{
\begin{array}{ll}
\frac{r}{r_0},  & r< r_0 \\
1 & r_0 < r< 1\\
r & r>1
\end{array}\right.  \quad 
K(r)=\left\{
\begin{array}{ll}
k_2 I   & r< r_0 \\
K_r=\infty, ~K_\theta =k_1\frac{r}{r_0 b} & r_0 < r< 1\\
k_* I  & r>1
\end{array}\right. 
$$
where $r_0 =\sqrt{m_2}$.

The energy is
$$ W= k_2 \int_0^{r_0}\left( \left( \frac{d \,u} {d\, r}\right)^2 + \frac{u^2}{r^2} \right)r\, dr + \int_{r_0}^1 K_\theta \frac{u^2}{r}  dr 
$$
or, substituting the value of $u(r)$,
$$ W=  k_2   + k_1 \frac{1-r_0}{b}=k_2+ k_1 \frac{\sqrt{m_2}(1-\sqrt{m_2})^2}{m_1}=B2
$$
This shows that $W(2, 13)$ assemblage is optimal for the intermediate regime. 
\paragraph{Transitional points}

when $m_1=m_{11}$, the expressions for bound B1 and B2 in \eq{b1}  coincide, and when   $m_1=m_{12}$, the expressions for B2 and B3 in \eq{b1}  coincide.

\subsection{Wheel assemblage $W(2, 13,1)$: Large $m_1$ (Hashin-Shtrikman bounds)}

In the interval $m_1> m_{11}$, the optimal symmetric structure $W(2, 13, 1) $ (Whole wheel, Figure \ref{f:wheel-large}) can be constructed as follows: circles of the Wheel Structure $W(2, 13) $ where $(m_1=m_{11})$ is enveloped by an annulus made from $k_1$. When $m_1 $ increases, the relative fractions in the central part stay constant, but the outer annulus becomes thicker.
\begin{figure}[htbp]
\begin{center}
\includegraphics[scale=.20]{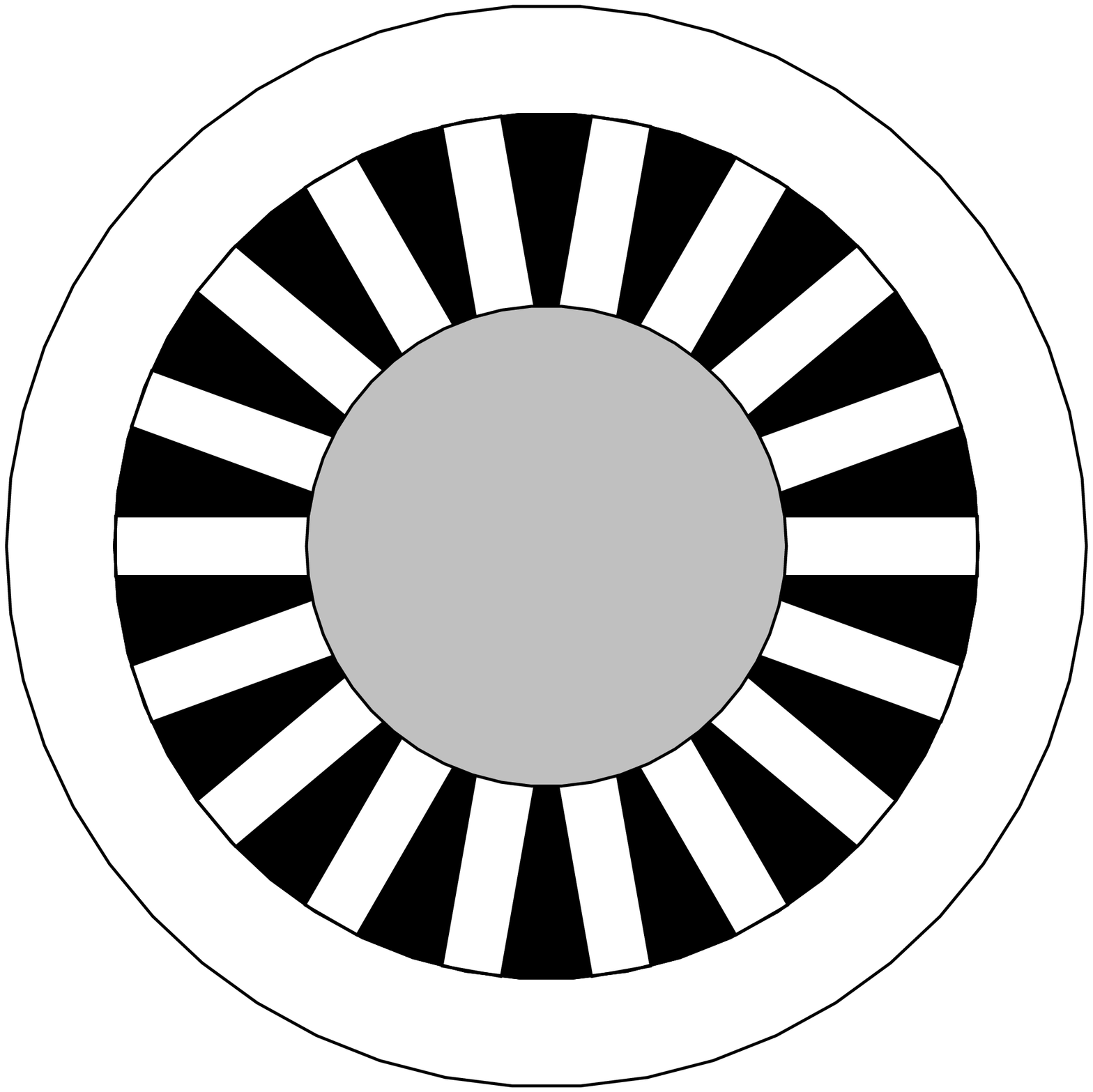} 
\caption{Cartoon of optimal laminates \cite{cherk09} that realize the bounds for isotropic three-material composite}
\label{f:wheel-large}
\end{center}
\end{figure}

To prove optimality of the stated structure, we notice  (see \cite{mybook,albin-nesi-cherk07}) that the effective conductivity $k_{cs}$ of a coated circles structure  satisfies the equation
\begin{equation}
\frac{1}{k_{cs}+k_1}= c \left( \frac{1}{2 k_1}-\frac{1}{k_{nucl}+k_1}\right) + \frac{1}{k_{nucl}+k_1}
\end{equation}
where $k_1$ is an enveloping material, $k_{nucl}$ is the core material, and $c$ is the added fraction of $k_1$. 
This implies that if a structure $k_{nucl}$ satisfies the Hashin-Shtrikman bound $B1$, the envelope of this structure by $k_1$ also satisfies it by virtue of algebraic form of the bound and coated circles formulas, see \cite{albin-nesi-cherk07,mybook}
One can check that the optimal $W(2, 13)$ structure satisfies both bounds $B_1$ and $B_2$ if $m_1$ is equal to the critical value $m_1=m_{11} $. Therefore,  $W(2, 13, 1)$ structure (Figure \ref{f:wheel-large}) satisfies $B1$ bound for all larger values of $m_1$.

\subsection{Wheel assembly $W(2,123)$. Small $m_1$}
 In the interval $m_1< m_{12}$, the optimal structures are
  wheel $W(2, 123)$, see Figure \ref{f:wheel-small}. 
\begin{figure}[htbp]
\begin{center}
\includegraphics[scale=.20]{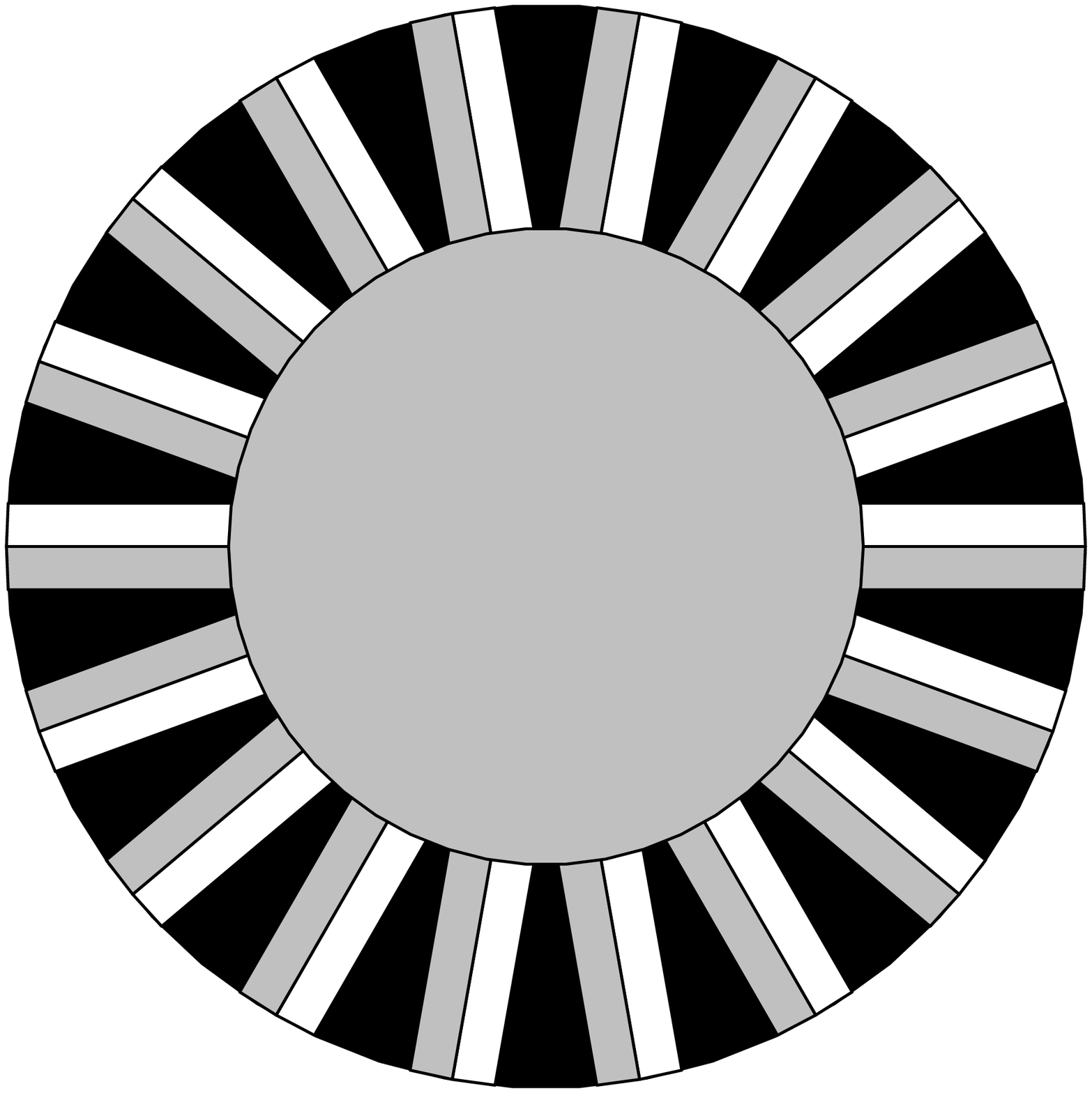} 
\caption{Cartoon of optimal wheel structures W(1, 123) that realize the bounds for isotropic three-material composite}
\label{f:wheel-small}
\end{center}
\end{figure}
These structures are similar to the basic wheel $W(2, 13)$  but the material $k_1$ in the spikes is replaced by a laminate of $k_1$ and $k_2$. At the point $r=r_0$, the volume fractions are
  $ b_1 $ and $b_2 $, respectively. 
Materials $k_1$ and $k_2$  occupy the areas
$$
m_1= 2 b_1 r_0 (1-r_0), \quad m_2=r_0^2+ 2 b_2 r_0 (1-r_0)
$$
and we find
\begin{eqnarray}
b_1 = \frac{m_1}{2 r_0 (1-r_0)}, \quad b_2=\frac{m_2-r_0^2}{2 r_0 (1-r_0)},
\end{eqnarray}

The continuity of potential at $r=r_0$ and the sufficient optimality conditions 
require the representation of the gradients at $r=r_0$
\begin{eqnarray}
e_{2c}=\frac{1}{r_0}\pmatrix{1 \cr 1} \quad  e_{2e}=\frac{1}{r_0}\pmatrix{2  \cr 0} \quad  e_{1e}=\frac{1}{r_0}\frac{ k_2}{k_1}\pmatrix{ 2\cr 0} 
\end{eqnarray}
Notice, that the continuity of current does not produce any constraints on fields because the current in superconductor phase $k_3$ is not defined. 

The optimal value $r_0$ of the radius of central patch filled with $k_2$ is computed as
$$ r_0=k_2 \left(\frac{m_1}{k_1}+\frac{m_2}{k_2}\right) $$
and corresponds to the energy equal to 
\begin{eqnarray}
W= k_2+ \left(  \frac{b_1}{k_1}+ \frac{b_2}{k_2}\right)^{-1}
\end{eqnarray}
that agrees with the bound $B3$. Condition $b_2\geq 0$ is equivalent to condition $m_1\leq m_{12}$.

\subsection{Asymptotics} Ii is interesting to find out how the three-material optimal structures degenerate into two-material ones.  The known optimal two-material structures are Coated Circles $W(2,1)$ when ($m_3\to 0$,  $W(3,2)$ when $m_1\to 0$, and  $W(3,1)$ when ($m_2\to 0$.  Asymptotics of three material wheel when $m_2  \ll 1$ and $m_3  \ll 1$ are  shown in Figure  \ref{equiv}, upper line. The structures degenerate into coated spheres. In the first case, the radius of the central sphere (equal to $\sqrt{m_2}$) goes to zero, and in the second � the thickness of the annulus goes to zero.
If $m_1=0$, the three-material wheel structure degenerates into $W(2, 23)$. The radius  $r_0$ of the
 inner circle filled with $k_2$ equals $m_2$, and the area fraction is $m_2^2$. 
 The remaining fraction $m_2(1-m_2)$  
of  $k_2$ is located in spikes, and $k_3$ in the trapezoidal areas between spikes. 
This structure is equivalent (it stored the same energy and has the same effective modulus $k_*$) to coated circles $W(3, 2)$ where $k_3$ is placed in the center circle and $k_2$ in the surrounding annulus, see Figure \ref{equiv}, lower line. The fields in these two structures are different but both satisfy sufficient conditions and the energy is the same. 

\begin{figure}[htbp]
\begin{center}
\includegraphics[scale=.15]{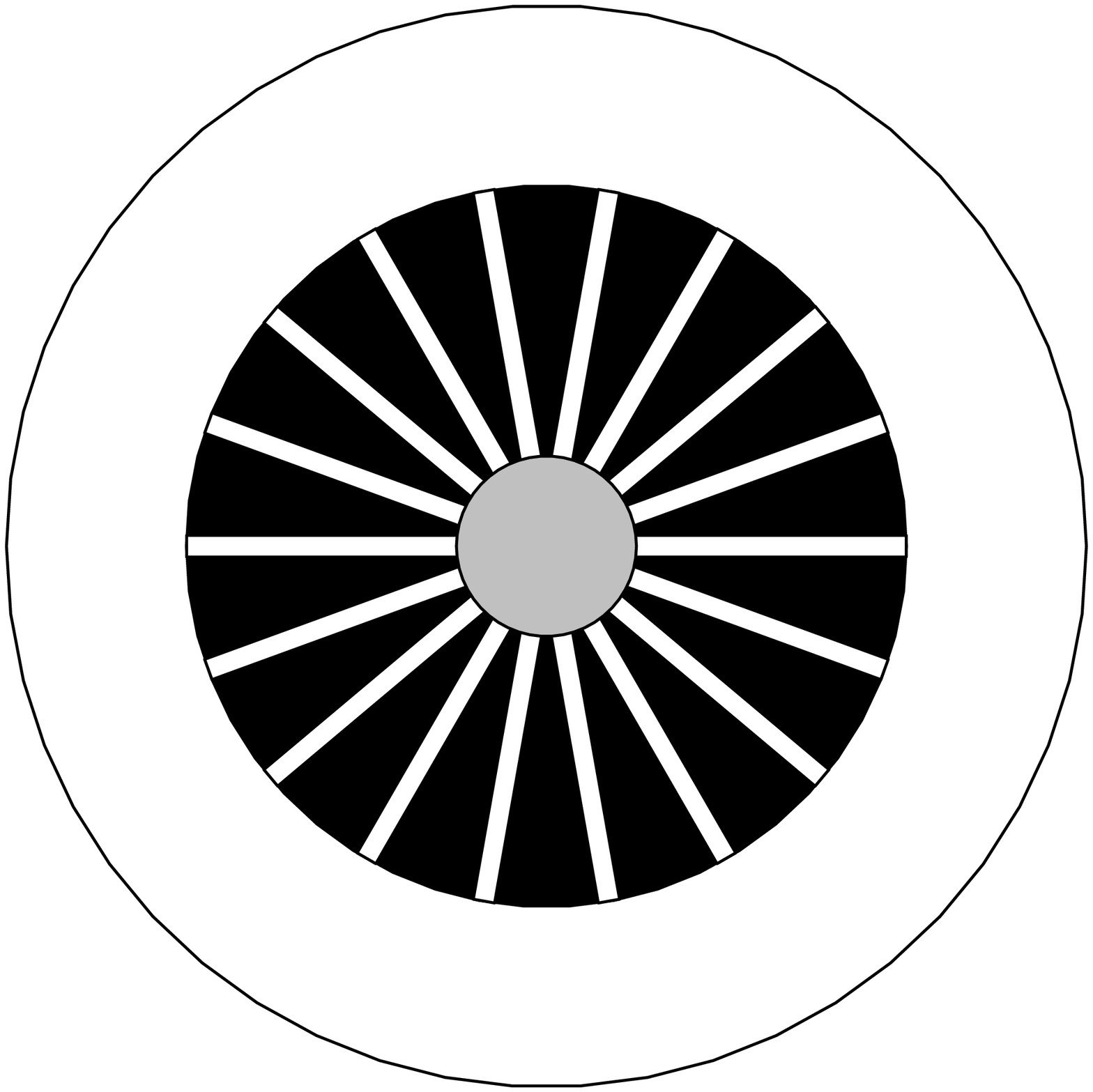} \qquad \includegraphics[scale=.15]{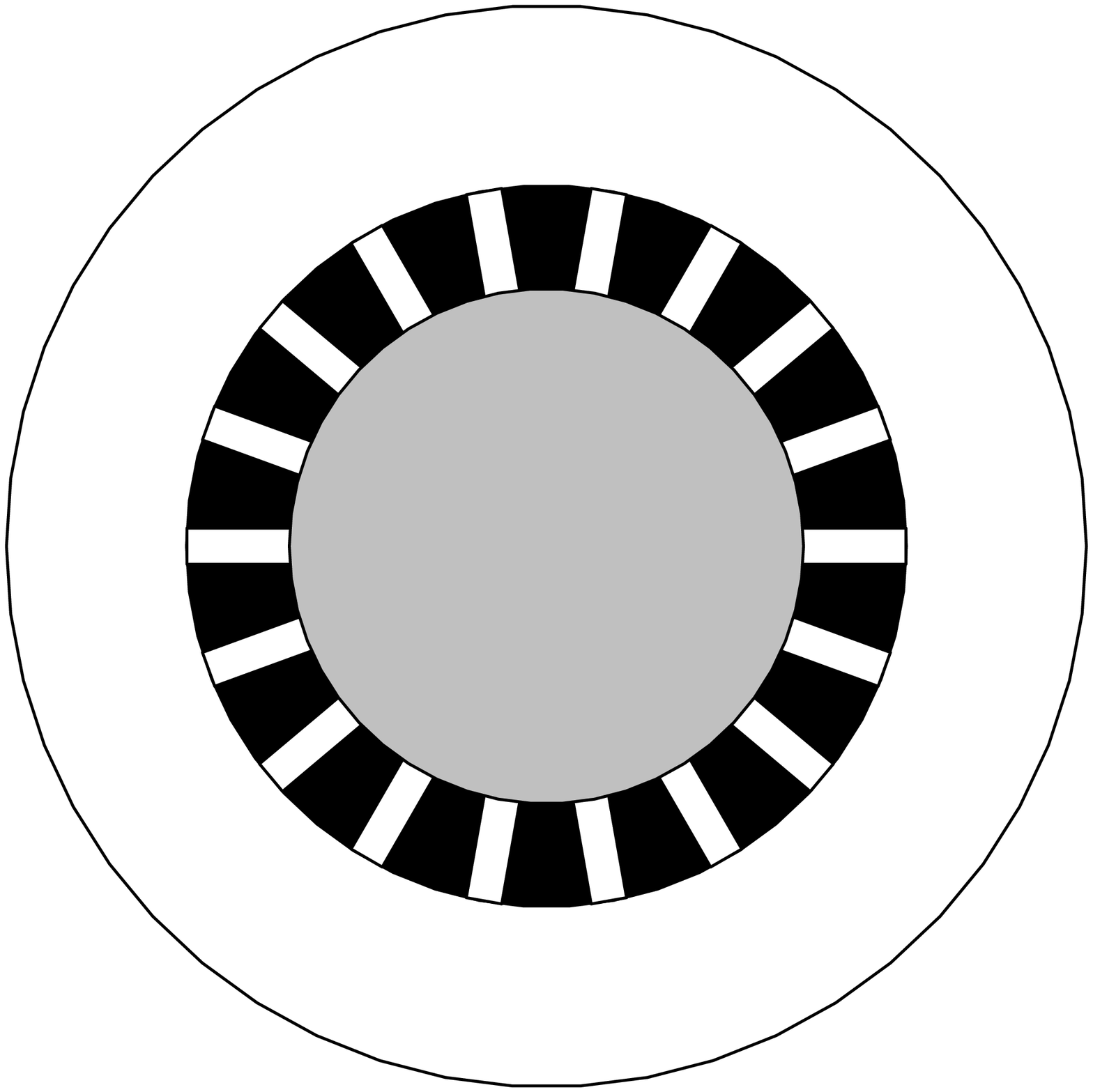}\\
~\\
\includegraphics[scale=.15]{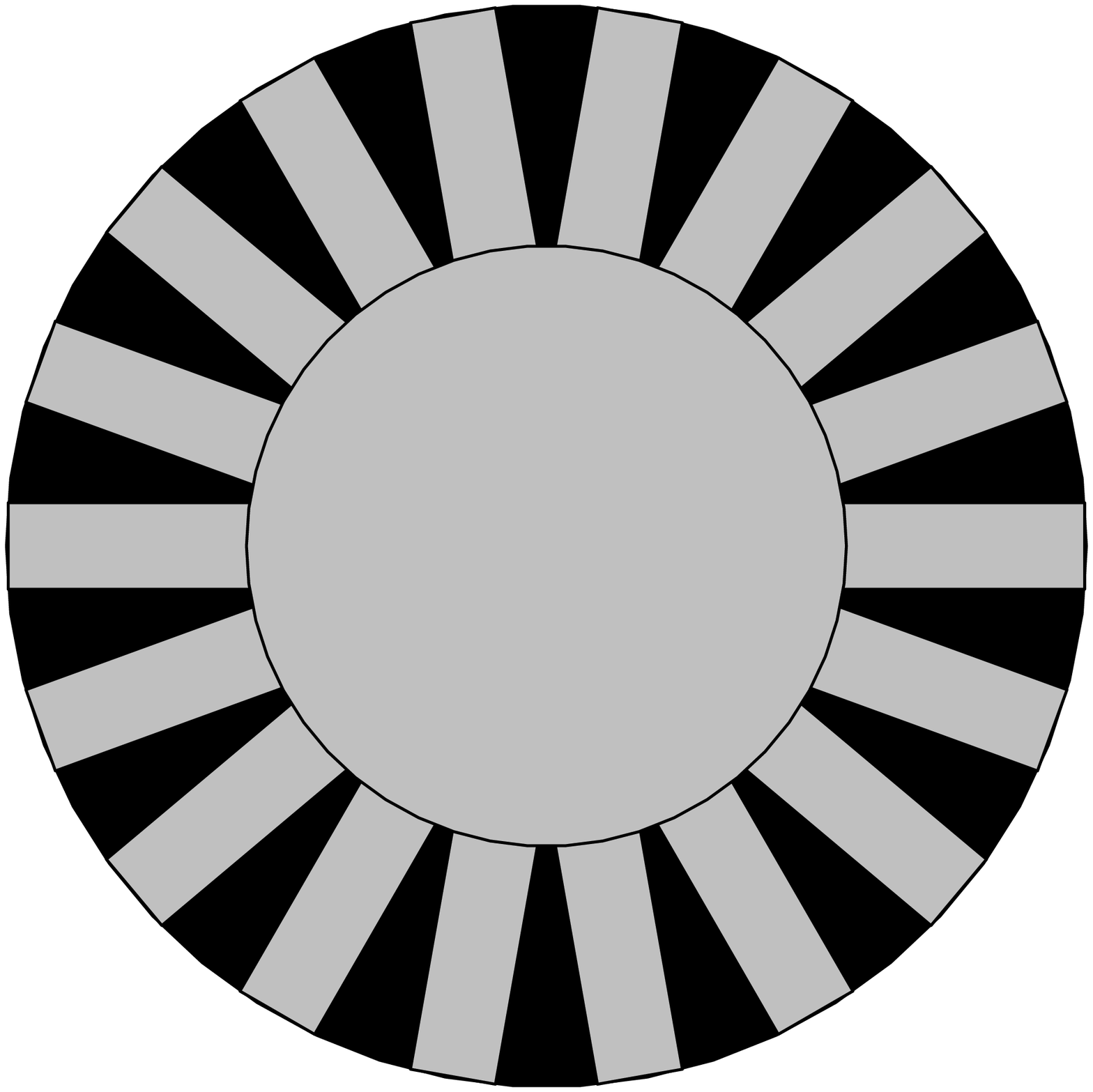}  \qquad \includegraphics[scale=.15]{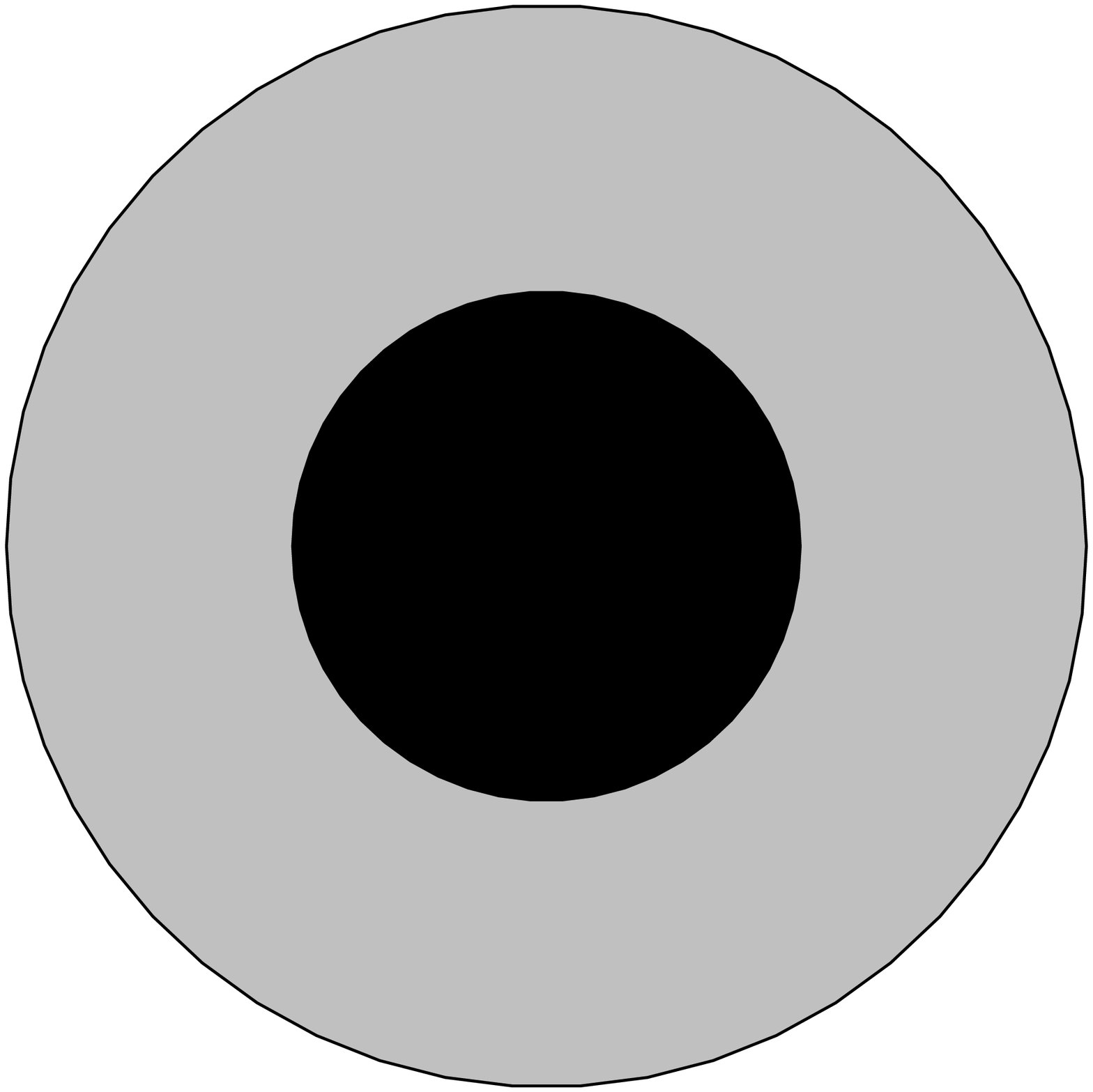}
\caption{
Upper Left: Optimal wheels, $m_2  \ll 1$,  Upper Right: Optimal wheels, $m_3 \ll 1$, 
Lower line: Limiting optimal wheel structure and equivalent coating circle structure: $m_1\to 0$. }
\label{equiv}
\end{center}
\end{figure}

\section{Dual problem. Two-material-and-void composites of minimal resistivity }

Structures similar to those described above also deliver minimal resistance of a composite made from two conductors and an insulator. Using the representation for any divergence-free vector $j$
\begin{equation}
j= R \nabla v, \quad R=\pmatrix{0 & 1 \cr -1 & 0}
\end{equation}
and accounting the identity $R^TR=I$, we present the energy of a material in the form
$$
W_j= \frac{1}{2} \frac{1}{k} |j|^2=  \frac{1}{2} \rho^2 |\nabla v |^2
$$
identical to \ref{en-cond}. Here, $ \rho=\frac{1}{k}$ is the resistance (a reciprocal to conductivity) of a material.
Because energy form is identical, the bounds \eq{b2} are valid where $k_i$ must be replaced b $\rho_i$, and it is assumed that $ \rho_1<\rho_2<\rho_3=\infty$.

\paragraph{Optimal Wheels structures of minimal resistivity}

The intermediate regime (bound B2, \eq{b2}) is realized by the Wheel $W(2,13)$ described above. This time, the trapezoidal segments filled with $\rho_3$ are insulators,  conductivity in the circumferential direction is zero and all current goes sequentially through the spikes and central circle.
For small $m_1$, bound B3 is realized by the Wheels structure $W(2, 123)$. 

Compare the currents in wheel $W(2, 123)$  in the primary problem with an ideal conductor in trapezoidal domains  and the same wheel in the dual problem with an ideal insulator in these domains. In the primary problem, the potential within each trapezoid is constant and the current in the spikes flows in circumferential directions across the layers, from one trapezoid to the next one. The current density is constant because layers have constant thickness. Another portion of currents flows through the central circle. The whole assembly is equivalent to a circuit: The layers in spikes are equivalent to sequentially jointed resistors, and the central circle is equivalent to a parallel resistor. In the dual problem, the trapezoids are filled with an ideal insulator and the circumferential current is zero. The current in spikes flows radially along the layers and then flows thought the central circle. The equivalent circuit consists of two parallel resistors that represent layers in the spikes, and a sequentially joined resistor that represents the central circle. These circuits are dual, because  the field in parallel resistors is constant as the current is in the sequential conductors, and vise versa.

Large values of $m_1$ correspond to the Hashin-Shtrikman bound and Wheels  $W(2, 13, 2)$. 
Notice that in each point of the same Wheel, the fields in the primary and dual problems are mutually 
orthogonal. In primary problem, the current in the spikes zone flows between the domain of superconductor 
across the layers, while in the dual one it flows along the layers parallel to domains of insulator. In the 
central circle, their currents are also orthogonal.

\subsection{Laminates and Wheels}

Let us compare the optimal laminate structure found in \cite{cherk09} and \cite{cz11}, shown in Figure \ref{f:lam}, with the wheel structures found here, shown in Figure \ref{f:allwheels}. The laminate structures \cite{cz11}  are anisotropic, and they depend on the degree of anisotropy and on volume fractions. Isotropic laminates are shown in the upper line. They also contain disconnected domains of $k_2$ where the field $E$ is proportional to a unit matrix, and $(k_1, k_3)$ or $(k_1, k_2, k_3)$ laminates that join these domains and direct currents between or around them. Larger amount of $k_1$ transforms the structure by placing the inclusions into a matrix from $k_1$; these structures realize the Hashin-Shtrikman bound. The wheel assemblages have similar features, but are simpler and easier to calculate. 

Comparing the wheel assemblage with anisotropic laminates for bounds B2 and B3, having small volume fractions 
of $m_1$ (the lower left and central fields in Figure \ref{f:lam}), we notice that the wheel structures
correspond
to these laminates if they are curled so that the lower layer becomes a circle and the upper layers become 
spikes and trapezoidal domains. The curled structure gains the axial symmetry and keeps the property 
of $(k_1, k_3)$ or $(k_1, k_2, k_3)$ laminates that direct currents between or around domains of $k_2$.

\begin{figure}[htbp]
\begin{center}
 \includegraphics[scale=.55]{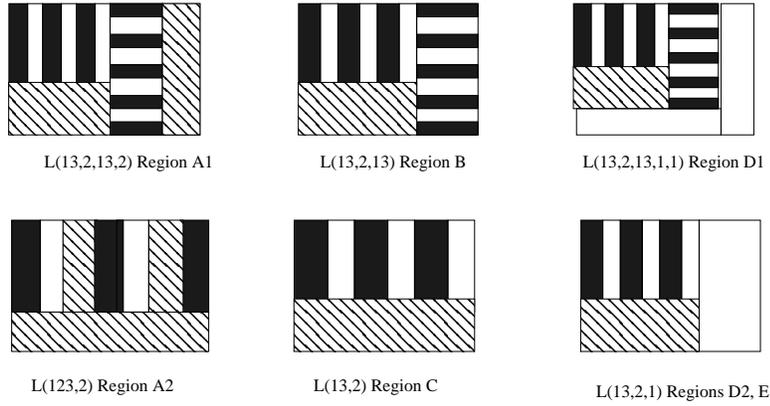} 
\caption{Cartoon of optimal laminates \cite{cherk09} that realize the bounds for isotropic three-material composite. They depend on the degree of anisotropy (vertical axis, top line corresponds to structures close to isotropy, bottom lime, to strongly anisotropic structures) and on the
volume fraction $m_1$ of the material $k_1$ (the volume fraction increases from left to right)}
\label{f:lam}
\end{center}
\end{figure}

\begin{figure}[htbp]
\begin{center}
B3. \includegraphics[scale=.14]{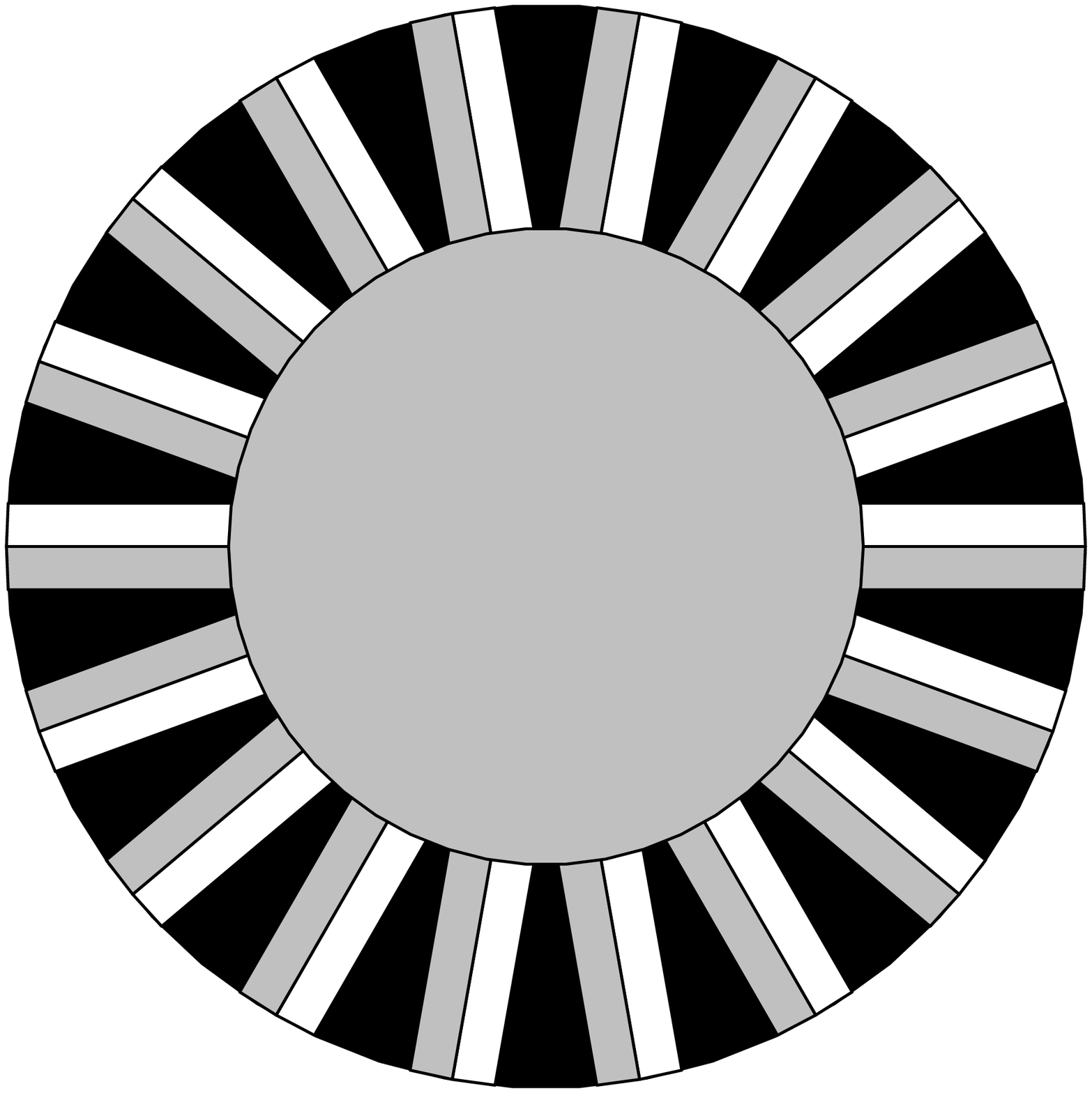} ~~
B2. \includegraphics[scale=.14]{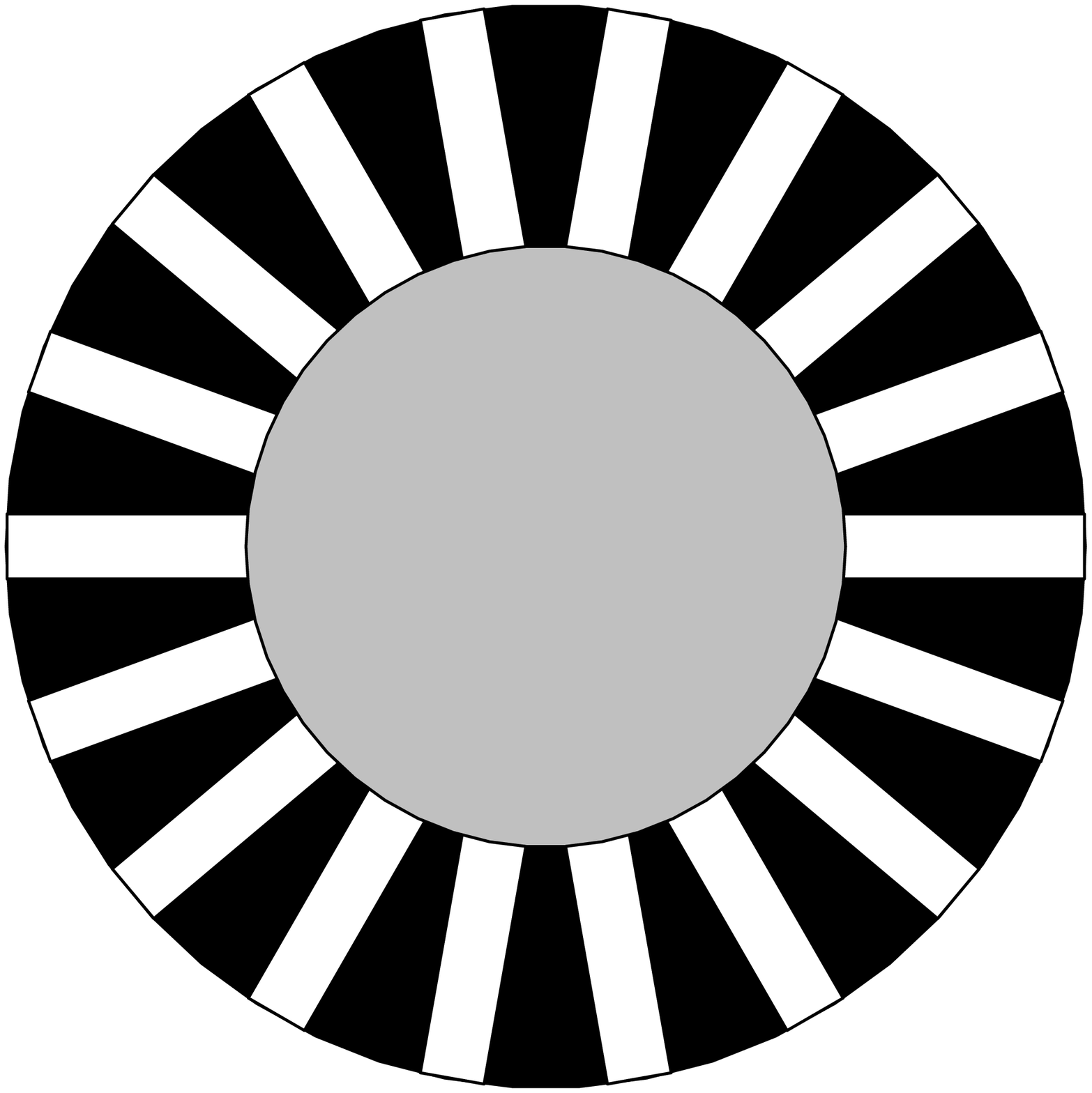} 
~~B1. \includegraphics[scale=.14]{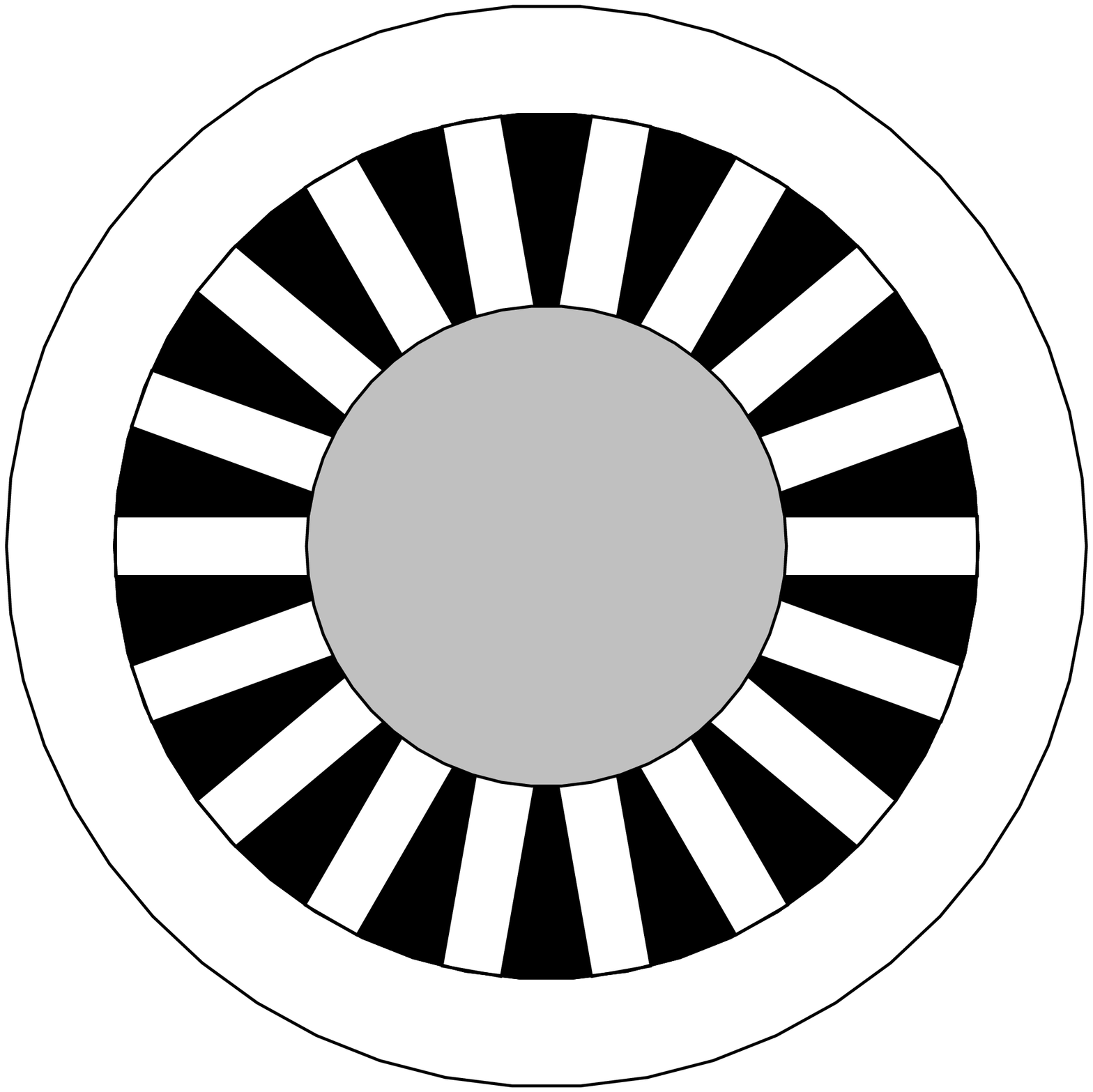} 

\caption{Cartoon of optimal Wheels structures that realize bound B1, B2, B3. }
\label{f:allwheels}
\end{center}
\end{figure}

A natural generalization of a wheel for anisotropic optimal composites would be a structure assembled from confocal ellipses instead of concentric circles and orthogonal hyperbolic layers instead of radial layers. Similar structures (excluding those with hyperbolic layers) are described in \cite{benveniste}.

\section{Elastic 2D structures from two materials and void with maximal bulk modulus}
The technique described above can be used to find the bound for the effective bulk modulus of a composite made from two linear elastic materials and void (plane problem). The elastic energy density $W_{el}$ is expressed through the stress tensor $\sigma$ as
\begin{equation}
W=\frac{1}{2} \left[\kappa ( \sigma_1)^2+ \eta ( \sigma_2)^2\right]
\label{en-elast}
\end{equation}
where $\sigma_1$ is the normalized trace and $\sigma_2$ is a magnitude of normalized Deviator (trace-free part)
of a $2\times 2 $ symmetric tensor $\sigma$:
\begin{equation}
\sigma_1=\frac{1}{\sqrt{2}} \mbox{Tr} \, \sigma= \frac{\sigma_{11}+ \sigma_{22}}{\sqrt{2}}, \quad  
\sigma_2=\sqrt{\frac{(\sigma_{11}- \sigma_{22})^2+ 2 \sigma_{12}^2 }{2}} 
\end{equation}
Notice that  
Tr\,$ \sigma^2 =2 \left(\sigma_1^2+ \sigma_2^2\right)$.
Here it is convenient to use elastic moduli $\kappa$ and $\eta$, the reciprocals of the plane bulk and shear moduli, respectively, and they are expressed through the Young's $E$ modulus and Poisson coefficient $\nu$ as
\begin{equation}
\kappa= \frac{1-\nu}{2 E} \quad  \eta=\frac{1+\nu}{2 E}.
\end{equation}
The moduli are assumed to be ordered as follows
\begin{equation}
\kappa_1 \leq \kappa_2 \leq \kappa_3= \infty, \quad \eta_1 \leq \eta_2 \leq \eta_3= \infty, 
\end{equation}

The Hooke's law corresponds to the Euler equation for energy \eq{en-elast} minimization and has the form
\begin{eqnarray*}
\varepsilon_{11}= (\kappa+ \eta) \sigma_{11} + (\kappa- \eta) \sigma_{22} \\
\varepsilon_{22}= (\kappa+ \eta) \sigma_{22} + (\kappa- \eta) \sigma_{11} \\
\varepsilon_{12}=  2 \eta \sigma_{12}  
\end{eqnarray*}
The equilibrium equations 
\begin{equation}
\nabla\cdot \sigma =0, \quad \sigma = \sigma^T
\end{equation}
allow for an Airy representation 
\begin{equation}
 \sigma =R^T (\nabla y )\, R, \quad  y=\nabla \chi .
\end{equation}
Because the rotated tensor $\sigma$ is a gradient of a vector $y$, we may reduce the problem to the previous case 
and establish the bounds for $\kappa$, in the same way it is done for two-material composites in \cite{mybook}. 
The lower bound of effective $\kappa$ modulus (that corresponds to the upper bound for effective bulk modulus)  is obtained similarly to the conductivity case using relation \eq{quasiaff} and \eq{nesi}.

Translated energies are
\begin{equation}
W^{\mbox{\footnotesize{transl}}}_1( s  ,  d , t) =\frac{1}{2}\left\{ 
\begin{array}{ll}
(\kappa+ t) (\sigma_1 )^2+ (\eta- t) \left( \sigma_2 \right) ^2 + \gamma_i &  \mbox{if } \det(\sigma) \geq 0 \\
+\infty & \mbox{if } \det(\sigma) < 0
\end{array}  
\right.
\end{equation}

When $ t\in [\eta_1, \eta_2] $, the energy  $F_1( s  ,  d , t)$ becomes nonconvex, its  convex envelope ${\cal C}F_1( s  ,  d  , t)$ is
\begin{equation}
{\cal C}F_1( s  , d  , t)=\left\{ 
\begin{array}{ll}
W^{\mbox{\footnotesize{transl}}}_1  & \mbox{if } \det(\sigma) \geq 0, ~0\le t\le \eta_1\\
\frac{\kappa_1 + \eta_1}{2}(\sigma_1)  ^2 + \gamma_1 & \mbox{if } \det(\sigma) \geq 0, ~t \geq  \eta_1\\
+\infty & \mbox{if } \det(\sigma) < 0
\end{array}  
\right.
\end{equation} 
In this case, the minimal energy corresponds to condition $\det \sigma=0$ in $\Omega_1$, corresponding to uniaxial stress in the first material.


To derive the bound, set the average stress field to be proportional to a unit tensor, ${(\sigma_0)}_1= 1$, ${(\sigma_0)}_2= 0$, and estimate the energy of a composite from below, obtaining bounds similar to \cite{nesi,cherk09}:
\begin{equation}
\kappa_* \geq B, \quad B=\max_{t\in[\eta_1, \eta_2]} \left[ -t + \left( \frac{m_1}{\kappa_1+\eta_1} +   \frac{m_2}{\kappa_2+t} \right)^{-1} \right]
\end{equation}
We compute:
\begin{itemize}
\item If $m_1>m_{11}$, then $t_{opt}=\eta_1$ and $B=B1$,
\begin{eqnarray}
B1= -\eta_1 + \left( \frac{m_1}{\kappa_1 + \eta_1} + \frac{m_2}{\kappa_2 + \eta_1}  \right)^{-1}
\label{elast-hs}
\end{eqnarray}
(Hashin-Shtrikman bound)

\item If $m_{12}< m_1<m_{11}$, then $\displaystyle{t_{opt}= \frac{(1-\sqrt{m_2}) (\kappa_1+\eta_1)}{m_1}-\eta_2}$ and $B=B2$,
\begin{eqnarray}
B2= \kappa_2 + \frac{(1-\sqrt{m_2})(\kappa_1+\eta_1)}{m_1}
\label{elast-interm}
\end{eqnarray}
(Intermediate bound)

\item If $m_1<m_{12}$, then $t_{opt}=\eta_2$ and $B=B3$,
\begin{eqnarray}
B3= -\eta_2 +   \left( \frac{m_1}{\kappa_1 + \eta_1} + \frac{m_2}{\kappa_2 + \eta_2}  \right)^{-1}
\label{elast-small}
 \end{eqnarray}
(small $m_1$ bound)
\end{itemize}
where the threshold values $m_{11}$ and $m_{12}$ of $m_1$ are:
\begin{eqnarray}
m_{11}=\sqrt{m_2} (1-\sqrt{m_2} )\frac{\kappa_1+\eta_1}{\kappa_2+\eta_1}, \quad 
m_{12}=\sqrt{m_2} (1-\sqrt{m_2} )\frac{\kappa_1+\eta_1}{\kappa_2+\eta_2}. 
 \end{eqnarray}

These bounds are exact. The optimal structures that realize the bounds are again the wheel assemblages described above. The calculation is similar to those shown above with obvious change in notations. Notice that the first term of the bound B2 \eq{elast-interm} 
corresponds to the bulk stress in the central circle, which is proportional to bulk modulus $\kappa_2$ of the material in the circle, and the second term corresponds to the stress in the spikes, which is inversely proportional to $\displaystyle{\kappa_1+\eta_1= \frac{1}{E_1}}$ the Young's modulus of the first material. 
This dependence is explained by observing that  the radial spikes are under one-directional radial load that they  transform to a  pressure on
the central circle. Similar structures realize the bound B3 \eq{elast-small}. For that case, the first term corresponds to the bulk stress in the center circle, and the second - to radial spikes. The second term also depends only on Young's moduli of the materials. The spikes in the annulus, Figures \ref{f:wheel-i}, \ref{f:wheel-small}  transport the hydrostatic stress to the central circle, providing optimal stiffness. The larger fraction of the stiffest material permits for additional outer annulus that increases the stiffness, Figure \ref{f:wheel-large}. 

\paragraph{Dual problem} The dual elasticity problem is addressed in the same manner. It asks for the minimal stiffness of three-material mixture, if one component is rigid. The bounds \eq{elast-hs}, \eq{elast-interm}, \eq{elast-small} are valid with a change in notations: $\kappa$ is changed to the plane bulk modulus $K$, $\displaystyle{K=\frac{1}{\kappa}}$ and $\eta $ is changed to shear modulus $\mu$, $\displaystyle{\mu=\frac{1}{\eta}}$.

The same wheel assemblages are optimal. We observe the dual behavior similar to the conductivity problem. The trapezoidal domains in an optimal wheel are filled with the rigid material which transmits the applied hydrostatic deformation to the middle circle and material in elastic spikes is under a uniaxial deformation  in the circumferential direction, the normal (circumferential) stress is constant.


\end{document}